\documentclass[amsmath,amssymb,aps,prd,onecolumn,groupedaddress,nofootinbib]{revtex4-2}

\usepackage{bm}
\usepackage{braket}
\usepackage{graphicx}
\usepackage{mathtools}
\usepackage{dcolumn}

\begin{document}
\title{Three-flavour neutrino oscillations in a magnetic field}

\author{Alexey Lichkunov}

\address{Department of Theoretical Physics, \\ Moscow State University, 119991 Moscow, Russia,\\ 
lichkunov.aa15@physics.msu.ru}

\author{Artem Popov}

\address{Department of Theoretical Physics, \\ Moscow State University, 119991 Moscow, Russia \\
	ar.popov@physics.msu.ru}

\author{Alexander Studenikin}

\address{Department of Particle Physics and Extreme States of Matter, \\ Department of Theoretical Physics, \\ Moscow State University, 119991 Moscow, Russia\\
	studenik@srd.sinp.msu.ru}

\begin{abstract}
In this paper we study Dirac and Majorana neutrino oscillations in a magnetic field in the three-flavour case. A theoretical framework developed in \cite{Popov:2019} is extended to the case of three neutrino species, as well as to the Majorana neutrinos case. The closed expression for the Dirac neutrinos flavour and spin oscillations are given. Majorana neutrino oscillations are studied numerically. We show that the probabilities exhibit a complicated interplay of oscillation on magnetic $\mu_\nu B$ and vacuum $\Delta m^2_{ij}/2p$ frequencies. It is also shown that neutrino oscillations in an interstellar magnetic field can modify neutrino fluxes observed in neutrino telescopes.
\end{abstract}

\maketitle


\section{Introduction}\label{sec:intro}

It is well-known that massive neutrinos have nontrivial electromagnetic properties (see \cite{Giunti:2015} for a review). In particular, neutrino magnetic moments are nonzero  \cite{Fujikawa, Schrock}. The best terrestrial upper bounds on neutrino magnetic moments on the level of $\mu_{\nu} < 2.9 \times 10^{-11} \mu_{B}$ were obtained by the GEMMA reactor neutrino experiment \cite{Beda} and $\mu_{\nu} < 6.4 \times 10^{-12} \mu_{B}$ by the XENON collaboration \cite{XENON:2022ltv} from  solar neutrino fluxes.

Interaction of neutrinos with nonzero magnetic moments with a transversal magnetic field leads to a phenomenon of \textit{neutrino spin precession} or \textit{neutrino spin oscillations}. Neutrino spin precession in the transversal magnetic field ${\bf B}_{\perp}$ was first considered in \cite{Cisneros}, spin-flavor precession was discussed in \cite{Schechter}. The resonant amplification of neutrino spin oscillations in ${\bf B}_{\perp}$ in the presence of matter was investigated in \cite{Akhmedov, Lim}.

In our previous paper \cite{Popov:2019} we proposed a new approach to the problem of neutrino mixing and oscillations in a magnetic field based on the exact solutions of the Dirac equation for a massive neutrino state in a magnetic field. Within the approach, a spin operator that commutes with the evolution Hamiltonian is used to classify the massive neutrino states in a magnetic field. We calculated the probabilities of neutrino flavour and spin oscillations in a magnetic field for the case of two neutrino flavours. We have shown that the probabilities exhibit an interplay of oscillations on the vacuum $\omega_{vac} = \Delta m^2/4p$ and magnetic $\omega_B = \mu_{\nu} B_{\perp}$ frequencies. In  \cite{Lichkunov:eps-hep2019,Lichkunov:ichep2020} we extended some of our results to the case of three neutrino flavours. The CP-violating effects in oscillations of Majorana neutrinos in a supernova media has been studied using our approach in \cite{Popov:2021}.

In this paper we continue our research of neutrino oscillations in a magnetic field and now apply our approach to the case of three neutrino flavours. The presented derivation of the explicit expressions for the probabilities of Dirac neutrino flavour and spin oscillations enable us to investigate the astrophysical neutrino flavour composition in an experiment for the three flavour case.

\section{Dirac neutrino flavour and spin oscillations in a magnetic field}

In this section we describe the approach to deriving of the Dirac neutrino oscillations probabilities in a magnetic field developed in \cite{Popov:2019} and extend it to the case of three neutrino flavours.

The wave function $\nu_i$ of a massive Dirac neutrino state that propagates in the presence of a constant and homogeneous and arbitrary orientated magnetic field  can be found as the solution of the following system of equations
\begin{equation}\label{eq1}
	(\gamma_{\mu} p^{\mu}-m_i-{\mu_i}{\bm{\Sigma}\bm{B}})\nu_i (p)=0,
\end{equation}
where $\mu_i$ are the neutrino magnetic moments, $i = {1,2,3}$. Note that in this paper we suppose that Dirac neutrinos do not possess the transition magnetic moments. Effects of the transition magnetic moments were studied in \cite{Kurashvili:2017} for the two flavour case.

Eq. \ref{eq1} can be rewritten in the equivalent Hamiltonian form
\begin{equation}
	H_i\nu_i= E_i \nu_i,
\end{equation}
where the Hamiltonian
\begin{equation}\label{Ham}
	H_i = m_i \gamma_0 + \gamma_0 \bm{\gamma}\bm{p} + \mu_i \gamma_0 \bm{\Sigma}\bm{B}.
\end{equation}
is introduced.

The eigenvalues of the Hamiltonian (\ref{Ham}) are given by

\begin{equation}\label{spec}
	E_i^s = \sqrt{m_i^2+p^2+{\mu_i}^2{B}^2 +2{\mu_i}s\sqrt{m_i^2{B}^2+p^2B_\bot^2}},
\end{equation}
where $s = \pm 1$, $p=|\bm{p}|$ and $B_\bot$ is a transversal (relatively to the neutrino momentum) component of the magnetic field.

Thus, stationary solutions of Eq. \ref{eq1} exist and satisfy the following relation
\begin{equation}\label{stat}
	H_i \ket{\nu_i^s(t)} = E_i^s \ket{\nu_i^s(t)}.
\end{equation}

To clarify the meaning of the spin number $s$ introduced in (\ref{spec}) and classify the neutrino stationary states in a magnetic field we use the following spin operator \cite{Popov:2019}
\begin{equation}\label{spin_oper}
	S_i = \frac{m_i}{\sqrt{m_i^2 \bm{B}^2 + \bm{p}^2 B^2_{\perp}}} \left[ \bm{\Sigma}\bm{B} - \frac{i}{m_i}\gamma_0 \gamma_5 [\bm{\Sigma}\times\bm{p}]\bm{B}\right].
\end{equation}

One can show that the spin operator (\ref{spin_oper}) commutes with the Hamiltonian
\begin{equation}
	[S_i,H_i]=0,
\end{equation}
and satisfy the following conditions
\begin{equation}
	S_i^2=1.
\end{equation}

Thus, the stationary states (\ref{stat}) are solutions of Eq. \ref{eq1} that are eigenstates of $S_i$:
\begin{equation} \label{spin_eigen}
	S_i \ket{\nu_i^s} = s \ket{\nu_i^s}, s = \pm 1,
\end{equation}
\begin{equation}\label{completeness}
	\sum_s \ket{\nu_i^s} \bra{\nu_i^s} = 1,
\end{equation}
\begin{equation}\label{normalizaton}
	\braket{\nu_i^s|\nu_k^{s'}} = \delta_{ik}\delta_{ss'}.
\end{equation}

In order to treat the evolution of the neutrino massive states in a magnetic field we expand the wavefunctions $\nu_i$ over the stationary states
\begin{eqnarray}\label{nu_c_d}
	\nu_i^L(t) = c_i^+ \nu_i^+(t) + c_i^- \nu_i^-(t) ,\\
	\nu_i^R(t) = d_i^+ \nu_i^+(t) + d_i^- \nu_i^-(t),
\end{eqnarray}
where $L,R$ are neutrino helicities,  $\nu_i^s(t) = \exp(-i E_i^s t)\nu_i^s(0)$, and the coefficients $c^{\pm}_i$ and $d^{\pm}_i$ do not depend on time and can be calculated using initial conditions.

We derive general expressions for the neutrino flavour and spin oscillations probabilities in a magnetic field using the common defition and strightforward expression
\begin{equation}
	P_{\nu_{\alpha}^h \rightarrow \nu_{\beta}^{h'}}(t) = \big| \braket{\nu_\beta^{h'}(0) | \nu_\alpha^h(t)} \big|^2 = \Big| \sum_{i} U^*_{\beta i} U_{\alpha i} \braket{\nu_i^{h'}(0)|\nu_i^h(t)}\Big|^2,
\end{equation}
where $U$ is the mixing matrix, $h,h'=L,R$ are helicities of neutrino initial and final states respectively, $\alpha, \beta = \{e,\mu,\tau\}$ are neutrino flavours and $i = \{1,2,3\}$. Using (\ref{nu_c_d}), we derive the following expressions for the probabilities of neutrino flavour and spin oscillations assuming that the initial state is an electron neutrino:

\begin{eqnarray}\label{pr1}
	P_{\nu_{e}^{L}\rightarrow \nu_{\alpha}^{L}}(t) =
	\Bigl| \left( |c_{1}^+|^2e^{-iE_{1}^+t}+|c_{1}^-|^2e^{-iE_{1}^-t} \right) U_{11} U_{\alpha1}^{*} + \nonumber \\ \left( |c_{2}^+|^2e^{-iE_{2}^+t}+|c_{2}^-|^2e^{-iE_{2}^-t} \right) U_{12}U_{\alpha2}^* + \\ \nonumber
	\left( |c_{3}^+|^2e^{-iE_{3}^+t}+|c_{3}^-|^2e^{-iE_{3}^-t} \right) U_{13}U_{\alpha3}^* \Bigr|^2,
\end{eqnarray}

\begin{eqnarray}\label{pr2}
	P_{\nu_{e}^{L}\rightarrow \nu_{\alpha}^{R}}(t) = \Bigl| \left( (d_{1}^+)^*c_{1}^+e^{-iE_{1}^+t}+(d_{1}^-)^*c_{1}^-e^{-iE_{1}^-t} \right) U_{11} U_{\alpha1}^{*} + \nonumber \\ \left( (d_{2}^+)^*c_{2}^+e^{-iE_{2}^+t}+(d_{2}^+)^*c_{2}^+e^{-iE_{2}^-t} \right) U_{12}U_{\alpha2}^* + \\ \nonumber \left( (d_{3}^+)^*c_{3}^+e^{-iE_{3}^+t}+(d_{3}^-)^*c_{3}^-e^{-iE_{3}^-t} \right) U_{13}U_{\alpha3}^* \Bigr|^2.
\end{eqnarray}

Below we calculate the coefficients $c_i^s$ and $d_i^s$ and generalise (\ref{pr1}) and (\ref{pr2}) to the case of an arbitrary flavour of initial neutrino.

Using (\ref{spin_eigen}), (\ref{completeness}) and (\ref{normalizaton}), one can show that
\begin{equation}
	\braket{\nu_i^{h'}(0)|\nu_i^h(t)} = \sum_s e^{-i E_i^s t} \braket{\nu_i^{h'}(0)|\nu_i^s}\braket{\nu_i^s|\nu_i^h(0)} = \sum_s e^{-i E_i^s t} \bra{\nu_i^{h'}(0)} \hat{P}_i^s \ket{\nu_i^h(0)},
\end{equation}
where the projection operators are introduced
\begin{equation}\label{projectors}
	P_i^{\pm} = \ket{\nu_i^{\pm}}\bra{\nu_i^{\pm}} = \frac{1 \pm S_i}{2}.
\end{equation}

Thus, the amplitudes of the transitions between massive neutrino helicity states are given by the plane wave expansion of the following form
\begin{equation}\label{planewave}
	\braket{\nu_i^{h'}(0)|\nu_i^h(t)} = \sum_s C_{is}^{h'h} e^{-i E_i^s t},
\end{equation}
where
\begin{equation}\label{decomp_coeffs}
	C_{is}^{h'h} = \bra{\nu_i^{h'}(0)} P_i^s \ket{\nu_i^h(0)}.
\end{equation}

The coefficients $C_{is}^{h'h}$ are related with $c_i^s$ and $d_i^s$ introduced in (\ref{nu_c_d}) by the following expressions
\begin{equation}
	|c_i^{s}|^2 = C_{is}^{LL}, \;
	|d_i^{s}|^2 = C_{is}^{RR}.
\end{equation}

Using the plane-wave expansion (\ref{planewave}) one can derive the general expression for the probabilities of neutrino oscillations for the case of three neutrinos as
\begin{equation}\label{prob_general}
	P_{\nu_{\alpha}^h \rightarrow \nu_{\beta}^{h'}}(t) = \sum_{i,j}\sum_{s,s'} U^*_{\beta i} U_{\alpha i} U_{\beta j} U_{\alpha j}^* C_{is}^{h'h} \left(C_{j s'}^{h'h}\right)^* e^{-i (E_i^s - E_j^{s'})t},
\end{equation}
where $s, s' = \pm 1$.

From (\ref{prob_general}) it can be shown that
\begin{equation}\label{prob_conservation}
	\sum_{\beta,h'} P_{\nu_{\alpha}^h \rightarrow \nu_{\beta}^{h'}}(t) = \sum_{\alpha,h} P_{\nu_{\alpha}^h \rightarrow \nu_{\beta}^{h'}}(t) = 1
\end{equation}
and
\begin{equation}
	P_{\nu_{\alpha}^h \rightarrow \nu_{\beta}^{h'}}(0) = \delta_{\alpha\beta}\delta_{hh'}.
\end{equation}

\section{Dirac neutrino oscillations probabilities in the ultrarelativistic limit}

The expression (\ref{prob_general}) for the probabilities of neutrino oscillations in a magnetic field can be significantly simplified for the ultrarelativistic case.

For the neutrino energy spectrum in the ultrarelativistic limit $m_i \ll p$  we get
\begin{equation}
	E^{s}_i \approx p + \frac{m^2_i}{2p} + \frac{\mu_i^2 B^2}{2p} + \mu_i s B_{\perp}.
\end{equation}
Given that neutrino magnetic moments are of order of $10^{-11} \mu_B$ or smaller, it is also reasonable to assume that $\mu_i B_{\perp} \ll m_i$ (see \cite{Popov:2019} for a detailed discussion). Thus we can approximately write neutrino energy spectrum in a magnetic field as
\begin{equation}\label{energy_s}
	E^{s}_i \approx p + \frac{m^2_i}{2p} + \mu_i s B_{\perp}.
\end{equation}

Since the differences of neutrino energies and the corresponding frequencies of neutrino oscillations are
\begin{equation}\label{freq}
	E^{s}_i-E^{s'}_j \approx \frac{\Delta m^2_{ij}}{2p} + (\mu_i s - \mu_j s')B_{\perp},
\end{equation}

we expect that the probabilities of neutrino oscillations in a magnetic field exhibit the inherent interplay of the oscillations on the vacuum frequencies $\Delta m^2_{ik}/4p$ and magnetic frequencies $\mu_i B_{\perp}$.

The plane-wave expansion coefficients $C^{hh'}_{is}$ also can be simplified if we account for the fact that in the ultrarelativistic limit neutrino helicity states are given by the corresponding limit of the Dirac equation solutions: 
\begin{equation}
	\ket{ \nu_i^L(0) } = \frac{1}{\sqrt{2}} \begin{pmatrix}
		0 \\
		-1 \\
		0 \\
		1
	\end{pmatrix}, \ \ \
	\ket{ \nu_i^R(0) } = \frac{1}{\sqrt{2}} \begin{pmatrix}
		1 \\
		0 \\
		1 \\
		0
	\end{pmatrix}.
\end{equation}

Then from (\ref{decomp_coeffs}) we have

\begin{equation}\label{LL}
	C^{LL}_{i s} \approx \frac{1}{2} \left( 1 + s \frac{m_i B_{\parallel}}{{\sqrt{m_i^2 B^2_{\parallel} + (m_i^2+p^2) B^2_{\perp}}}} \right),
\end{equation}
\begin{equation}\label{RL}
	C^{RL}_{i s} \approx - \frac{s}{2}\frac{p B_{\perp}}{\sqrt{m_i^2 B^2_{\parallel} + (m_i^2+p^2) B^2_{\perp}}}.
\end{equation}
In the derivation above we use the fixed reference frame for which $B_{\parallel} = B_z$, $\bm{B_{\bot}} = (B_x,0,0)$. Note that in the limit $B_{\bot} = 0$ the coefficients $C^{RL}_{i s} = 0$, and the probabilities of neutrino spin oscillations $\nu_\alpha^L \to \nu_\beta^R$ are zeros. Thus, neutrino spin oscillations are generated by the transversal magnetic field, as it should be.

Finally, in the ultrarelativistic limit we assume that $t \approx x$, where $x$ is the distance travelled by neutrino.

Neglecting term of order of $m_i^2/p^2$ and smaller, we arrive to the expressions for the neutrino flavour $\nu^L_\alpha \to \nu^L_\beta$ and spin $\nu^L_\alpha \to \nu^R_\beta$ oscillations probabilities:
\begin{equation}\label{prob_fl}
	P_{\nu_{\alpha}^L \to \nu_{\beta}^L}(x) = \frac{1}{4} \sum_{i,j} U^*_{\beta i} U_{\alpha i} U_{\beta j} U_{\alpha j}^* \sum_{s,s'} e^{-i (E_i^s - E_j^{s'})x},
\end{equation}
\begin{equation}\label{prob_sf}
	P_{\nu_{\alpha}^L \to \nu^R_\beta}(x) = \frac{1}{4} \sum_{i,j} U^*_{\beta i} U_{\alpha i} U_{\beta j} U_{\alpha j}^* \sum_{s,s'} ss' e^{-i (E_i^s - E_j^{s'})x}.
\end{equation}

These expressions can be further simplified. For the probabilities of neutrino three-flavour oscillations we get
\begin{eqnarray}\label{prob_fl_final}
	&P_{\nu_{\alpha}^L \to \nu_{\beta}^L}(x) = \sum_{i=1}^3 |U_{\alpha i}|^2 |U_{\beta i}|^2 \cos^2( \mu_i B_{\perp}x ) + \sum_{i>j} 2\cos(\mu_i B_{\perp}x) \cos(\mu_j B_{\perp}x) \times \nonumber \\
	&\times \left[ \text{Re}(A_{ij}^{\alpha \beta}) \cos\Big( \frac{\Delta m^2_{ij}}{2p}x \Big) + \text{Im}(A_{ij}^{\alpha \beta}) \sin\Big( \frac{\Delta m^2_{ij}}{2p}x \Big) \right],
\end{eqnarray}
where $A_{ij}^{\alpha \beta} = U^*_{\beta i} U_{\alpha i} U_{\beta j} U_{\alpha j}^*$. Note that $\text{Im} A^{\alpha\beta}_{ij} = J \sin \delta \sum_{k,\gamma} \epsilon_{\alpha\beta\gamma} \epsilon_{ijk}$, where $\delta$ is CP-violating phase and $J \approx 0.034$ is the leptonic Jarlskog invariant (see \cite{Esteban:2020}). Thus, the last term in (\ref{prob_fl_final}) describes CP-violating effects.

It makes sense to introduce a new observable that describes the total probability of conversion into a sterile neutrino state $P_{\nu_{\alpha}^L \to \nu^R} = P_{\nu_{\alpha}^L \to \nu^R_e} + P_{\nu_{\alpha}^L \to \nu^R_\mu} + P_{\nu_{\alpha}^L \to \nu^R_\tau}$:
\begin{equation}
	P_{\nu_{\alpha}^L \to \nu^R}(x) = \frac{1}{4} \sum_{\beta} \sum_{i,j} U^*_{\beta i} U_{\alpha i} U_{\beta j} U_{\alpha j}^* \sum_{s,s'} ss' e^{-i (E_i^s - E_j^{s'})x}.
\end{equation}
This expression can be further simplified. Using the unitarity condition of the mixing matrix $\sum_{\beta}  U^*_{\beta i} U_{\beta j} = \delta_{ij}$ and the fact that $E_i^L - E_i^{R} = 2 \mu_i B_{\perp}$, we arrive to the final expression for the probability of neutrino spin oscillations

\begin{equation}\label{spin_total}
	P_{\nu_{\alpha}^L \to \nu^R}(x) = \sum_{i=1}^{3} |U_{\alpha i}|^2 \sin^2(\mu_i B_{\perp}x)
\end{equation}

Note that the probability of spin oscillations (\ref{spin_total}) does not depend on the value of the CP-violating phase since it depends only on absolute values of the mixing matrix entries.

\section{Majorana neutrino oscillations in a magnetic field}
As neutral fermions, neutrinos can be not only Dirac, but Majorana particles. 
Majorana neutrino interaction with a magnetic field $\bm{B}$ is described (see \cite{Popov:2021,Popov:2023wif,Lichkunov:2025rpu}) by the following Lagrangian
\begin{eqnarray}
\label{mag_field_int}
\mathcal{L}_{mag} = -\sum_{ik}\mu_{ik}\left[ \overline{(\nu_i^L)^c} \bm{\Sigma}\bm{B} \nu_k^L + \overline{\nu_i^L} \bm{\Sigma}\bm{B} (\nu_k^L)^c \right] = \\ \nonumber
 \sum_{\alpha\beta}\left[(\mu^{(f)})^\dag_{\alpha\beta} \overline{\nu_{\alpha}^L} \bm{\Sigma}\bm{B} (\nu_{\beta}^L)^c - \mu_{\alpha \beta}^{(f)} \overline{(\nu_{\alpha}^L)^c} \bm{\Sigma}\bm{B} \nu_{\beta}^L \right],
\end{eqnarray}
where $\mu$ is the neutrino magnetic moments matrix of the mass neutrino states and 

\begin{equation}\label{mm_flavour}
\mu^{(f)}=U\mu U^T,
\end{equation}
is the Majorana neutrino magnetic moments matrix in the flavour basis. The mixing matrix $U$ for the case of Majorana neutrinos is given by

\begin{equation}\label{PMNS}
	U=\begin{pmatrix}
	1 & 0 & 0 \\
	0 & c_{23} & s_{23} \\
	0 & -s_{23} & c_{23}
	\end{pmatrix}
	\begin{pmatrix}
	c_{13} & 0 & s_{13} e^{-i\delta} \\
	0 & 1 & 0 \\
	-s_{13} e^{i\delta} & 0 & c_{13}
	\end{pmatrix}
	\begin{pmatrix}
	c_{12} & s_{12} & 0 \\
	-s_{12} & c_{12} & 0 \\
	0 & 0 & 1
	\end{pmatrix}
	\begin{pmatrix}
	e^{i\alpha_1} & 0 & 0 \\
	0 & e^{i\alpha_2} & 0 \\
	0 & 0 & 1
	\end{pmatrix},
\end{equation}
where $\delta$ is the Dirac CP-violating phase and $\alpha_1$ and $\alpha_2$ the Majorana CP-violating phases. Note that the magnetic moments matrix in the flavour basis $\mu^{(f)}$ given by (\ref{mm_flavour}) depends on the both Dirac and Majorana CP-violating phases values.

Unlike in the case of Dirac neutrinos, the interaction Lagrangian (\ref{mag_field_int}) describes transitions between neutrinos $\nu_\alpha$ and antineutrinos $(\nu_\beta)^c$ of different flavours $\alpha$ and $\beta$.

It can be shown that as a consequence of CPT-invariance and hermiticity of the interaction Lagrangian (\ref{mag_field_int}), the Majorana neutrinos magnetic moments matrix in the mass basis is antisymmetric and imaginary, and can be parametrized as follows \cite{Giunti:2015}

\begin{equation}
\mu = ||\mu_{ij}||=\begin{pmatrix}
0 & i\mu_{12} & i|\mu_{13}| \\
-i|\mu_{12}| & 0 & i|\mu_{23}| \\
-i|\mu_{13}| & -i|\mu_{23}| & 0
\end{pmatrix}.
\end{equation}

The Majorana neutrinos wave functions are the solutions of the following Dirac equation that is derived using the Lagrangian (\ref{mag_field_int})
\begin{equation}\label{dirac_eqn_majorana}
    ( \gamma_\mu p^\mu - m_i )\nu_i(p) + \sum_{k\neq i} \mu_{ik}\bm{\Sigma}\bm{B}\nu_k(p)= 0.
\end{equation}
Unfortunately, Eq. (\ref{dirac_eqn_majorana}) has compact analytical solution only in the case of two neutrino flavours \cite{Kurashvili:2017,Lichkunov:2025rpu}. In this paper we solve the evolution equation (\ref{dirac_eqn_majorana}) numerically and compute the neutrino-neutrino and neutrino-antineutrino oscillations probabilities using the approach described in our papers \cite{Popov:2021,Popov:2023wif}.

\section{Ultra-high energy neutrino oscillations in interstellar media}\label{sec:v}

As an example of neutrino evolution in a magnetic field we study propagation of ultra-high energy neutrinos in the interstellar media (the case of two neutrino flavours was studied in \cite{Kurashvili:2017}). 
As an example we consider cosmogenic neutrinos that originate from interaction of ultra-high energy cosmic rays with the cosmic microwave background \cite{Berezinsky:1969erk}. Cosmogenic neutrino fluxes typically have energies of order of $1$ EeV (i.e. $10^{18}$ eV) \cite{Aloisio:2015ega}. In addition, theoretically proposed mechanisms exist that predict production of neutrinos with even higher energies. For example, cosmic strings are considered as a source of neutrinos with energies up to $10^{25}$ eV \cite{Berezinsky_strings}. A review of the possibility of ultra high-energy detection in future neutrino experiments can be found in \cite{Ackermann:2022rqc}.

We start with the case of Dirac neutrinos. Consider neutrinos with energy $p = 1$ EeV and , $B = 2.8$ $\mu$G for the interstellar magnetic field strength (see \cite{Zirnstein1:2016, Grasso:2001}). We also set the value of CP-violating phase $\delta$ to zero. For the sake of simplicity we start our analysis with the case of equal neutrino magnetic moments: $\mu_1 = \mu_2 = \mu_3 = 6.4\times10^{-12} \mu_B$ \cite{{Popov:2023wif}}. The probabilities of neutrino flavour oscillations $\nu_e \to \nu_{e}$ , $\nu_e \to \nu_{\mu}$ and $\nu_e \to \nu_{\tau}$ as functions of distance traveled by neutrino in parsec are shown in Fig. 1a and Fig. 1b. The probabilities indeed exhibit a complicated interplay of oscillations on the following vacuum and magnetic lengths:

\begin{equation}\label{lengths_B}
	L^B_i = \frac{\pi}{\mu_i B_{\perp}} =  3.49 \cdot \left( \frac{B}{\text{G}}\right)^{-1} \left( \frac{\mu_i}{10^{-12}\mu_B} \right)^{-1} 10^{-3} \; \text{pc},
\end{equation}

\begin{equation}\label{lengths_vac}
	L^{vac}_{ij} = \frac{4 \pi p}{\Delta m^2_{ij}} = 8.07 \cdot \left( \frac{\Delta m_{ij}^2 }{\text{eV}^2}\right)^{-1} \left( \frac{p}{\text{GeV}} \right) \cdot 10^{-14} \; \text{pc}.
\end{equation}

For the choice of parameters given above, oscillations on the vacuum lengths $L^{vac}_{12} \approx 1.08$ pc and $L^{vac}_{13} \approx 0.032$ pc that can be observed at smaller scales (Fig. 1a) are modulated by oscillations on the magnetic length $L^{B} \approx 195$ pc (Fig. 2a). This behaviour reproduces phenomenon observed in \cite{Kurashvili:2017, Popov:2019} for the case of two neutrino flavours. For the case of $\mu_1 = \mu_2 = \mu_3$,  Eq. (\ref{prob_fl_final}) reduces to

\begin{equation}\label{prob_fl}
	P_{\nu_{\alpha}^L \to \nu_{\beta}^L}(x) = \cos^2( \mu B_{\perp}x ) \left[ \sum_{i=1}^3 |U_{\alpha i}|^2 |U_{\beta i}|^2 + 2\sum_{i>j}  A_{ij}^{\alpha \beta} \cos\Big( \frac{\Delta m^2_{ij}}{2p}x \Big) \right],
\end{equation}
Here the expression in the square brackets is the vacuum oscillations probability and it is modulated by oscillations on magnetic frequency $\mu B_{\perp}$.

The interplay of oscillations on the vacuum and magnetic frequencies becomes even more pronounced if neutrino magnetic moments are not equal to each other. In Fig. 2a we show the probabilities of neutrino flavour oscillations for this particular case  

In Fig. 2b we plot the neutrino spin oscillations probability $\nu_e \to \nu^R$ for different values of the neutrino magnetic moment $\mu_1$. The magnetic moment $\mu_2$ value is fixed as $6.4\times10^{-12} \mu_B$. Since the last term in the sum in (\ref{spin_total}) for the case of initial electron is proportional to a quite small value $\sin^2 \theta_{13} \approx 0.022$, we can neglect the contribution of the magnetic moment $\mu_3$. In the case of equal magnetic moments $\mu_1 = \mu_2 = \mu$ neutrino spin oscillations probability is 
\begin{equation}
	P_{\nu_e \to \nu^R}(x) = \sin^2(\pi x/L_B),
\end{equation}
where the oscillations length $L_B \approx 195$ pc. For different choices of the neutrino magnetic moments oscillations probability exhibit an interplay of oscillations on two magnetic frequencies $\omega_1 = \mu_1 B_{\perp}$ and $\omega_2 = \mu_2 B_{\perp}$. Note that distance-averaged spin oscillations probability equals $1/2$ regardless of choice of neutrino magnetic moments.

\begin{figure}[h]
	\begin{minipage}[h]{0.49\linewidth}
		\center{\includegraphics[width=1\linewidth]{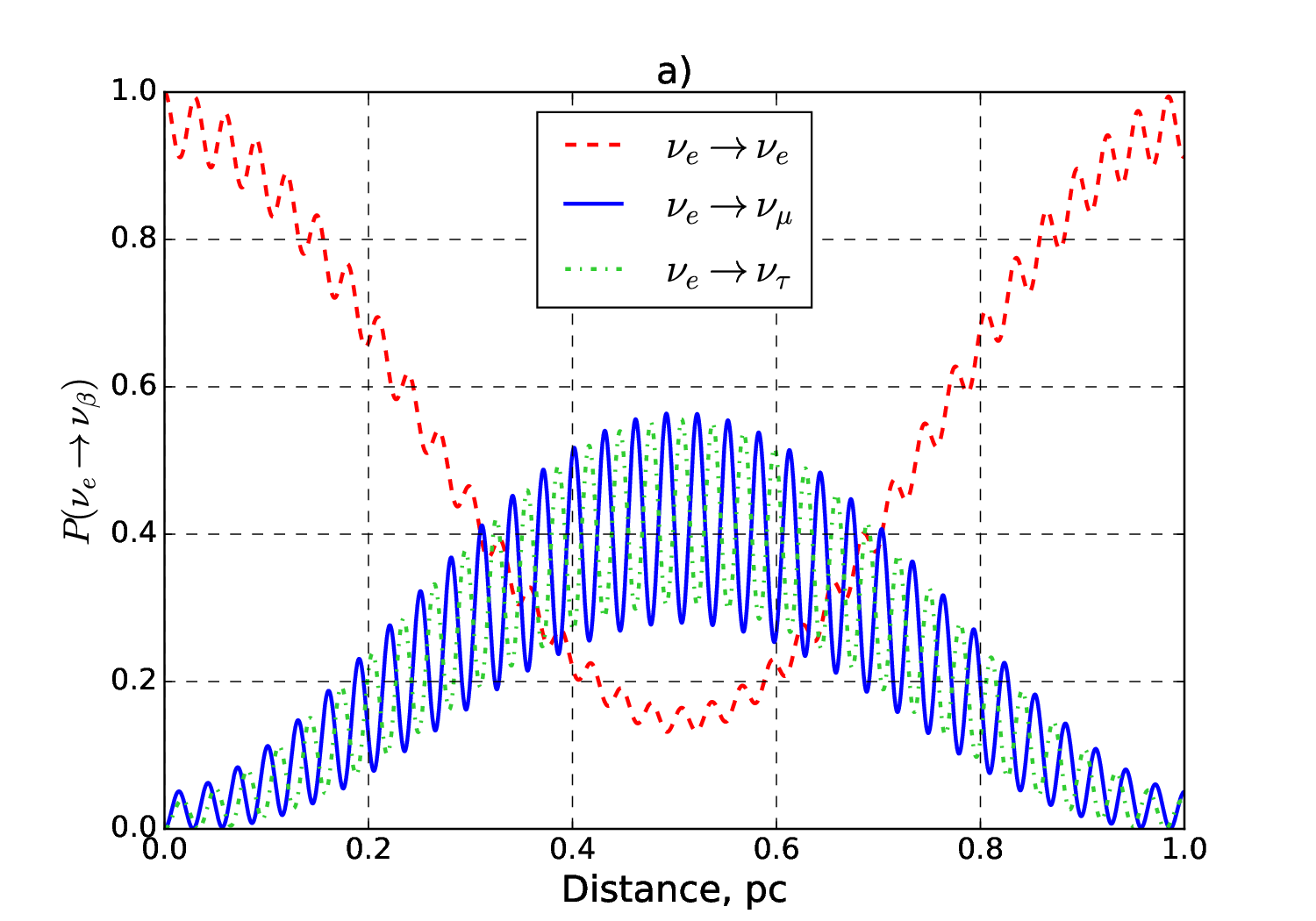}}
	\end{minipage}
	\begin{minipage}[h]{0.49\linewidth}
		\center{\includegraphics[width=1\linewidth]{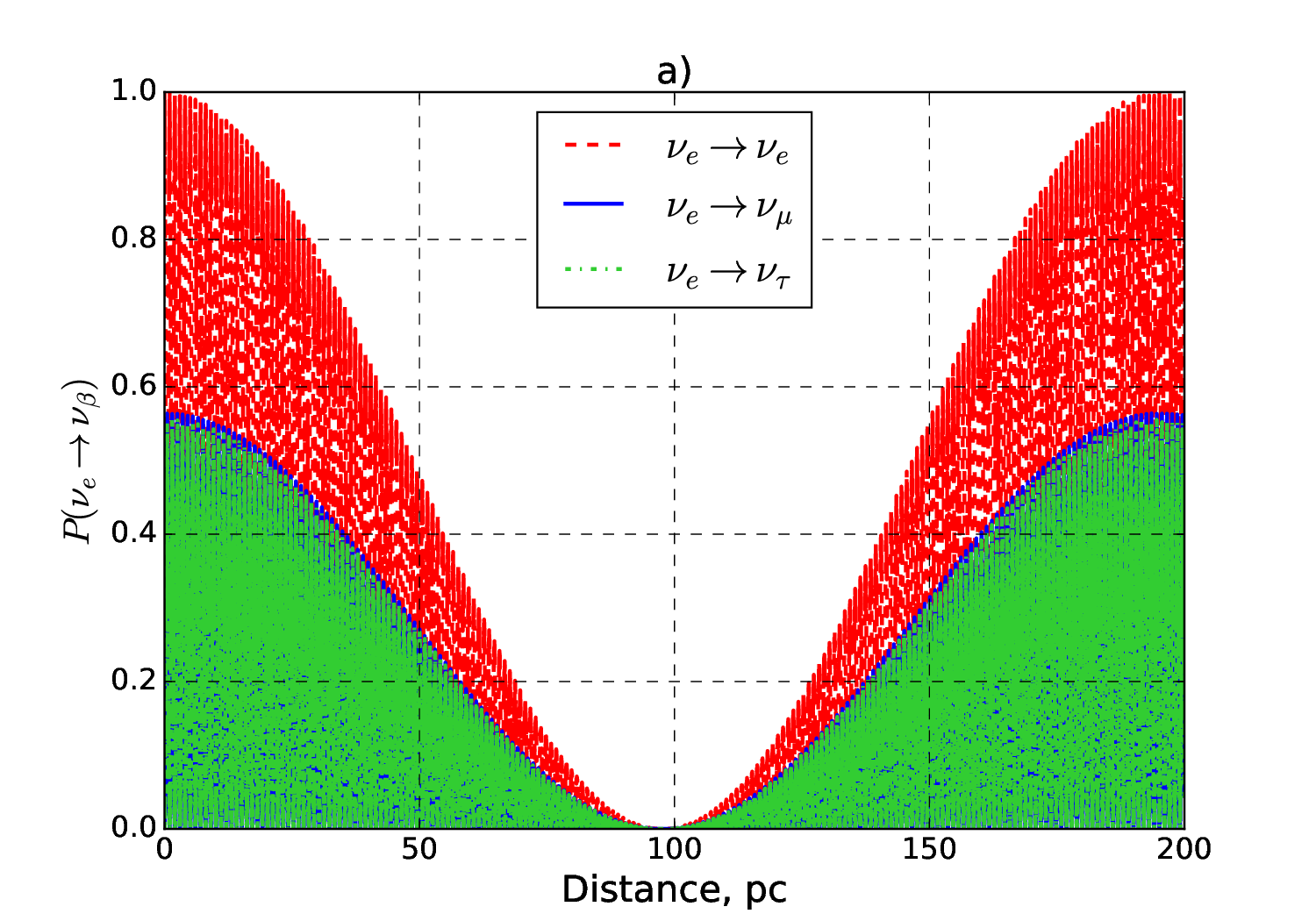}}
	\end{minipage}
	\vspace{-3mm}
	\caption{Dirac neutrino flavour oscillations probabilities for neutrino energy $E = 1$ EeV and magnetic moments $\mu_1=\mu_2=\mu_3=6.4\times10^{-12} \mu_B$.}
	\vspace{-5mm}
\end{figure}

\begin{figure}[h]
	\begin{minipage}[h]{0.49\linewidth}
		\center{\includegraphics[width=1\linewidth]{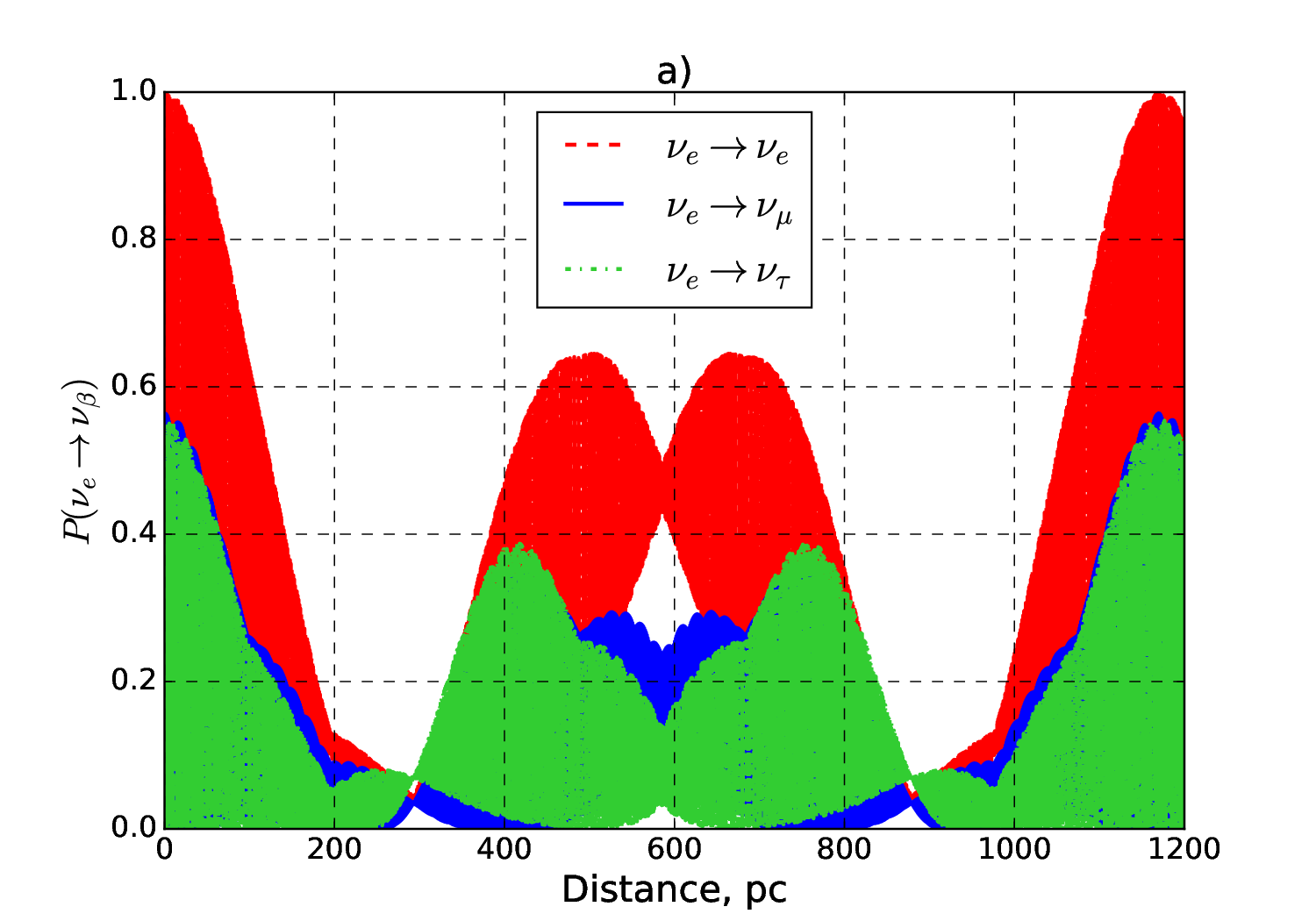}}
	\end{minipage}
	\begin{minipage}[h]{0.49\linewidth}
		\center{\includegraphics[width=1\linewidth]{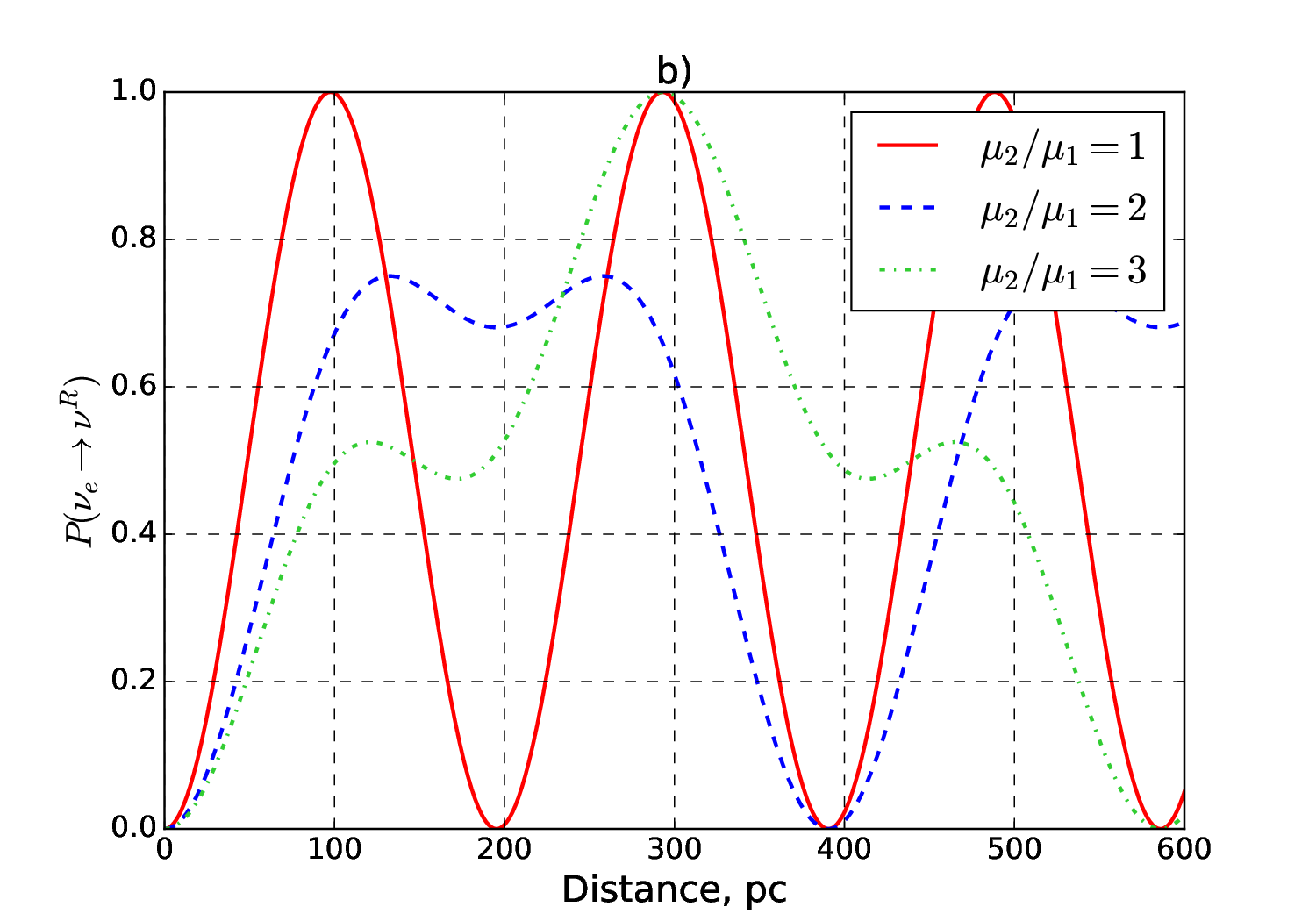}}
	\end{minipage}
	\vspace{-3mm}
	\caption{a) Dirac neutrino flavour oscillations probabilities for neutrino energy $E = 1$ EeV and magnetic moments $\mu_1=6.4/3\times10^{-12} \mu_B$, $\mu_2=6.4/2\times 10^{-12} \mu_B$ and $\mu_3=6.4\times 10^{-12} \mu_B$; b) Dirac neutrino spin oscillations probability $\nu_e^L \to \nu^R$ for different values of magnetic moments assuming $\mu_1 = 6.4\times 10^{-12} \mu_B$.}
	\vspace{-1mm}
\end{figure}

As was mentioned above, in the case of Majorana neutrinos interaction with a magnetic field induces transitions between neutrinos and antineutrinos of different flavours. However, this effect appears only for extremely high neutrino energies and/or magnitudes of a magnetic field.

In Fig.3 we show the probabilities of neutrino-neutrino and neutrino-antineutrino oscillations in the interstellar magnetic field assuming that the transition magnetic moments are given by $|\mu_{12}|=|\mu_{13}|=|\mu_{23}| = 6.4\cdot10^{-12} \mu_B$. We consider neutrino energies $1$ ZeV = $10^{21}$ eV, since for lower energies the effects due to interaction with the magnetic field disappear. As in the case of Dirac neutrinos, the oscillations probabilities exhibit a complicated interplay on the magnetic and vacuum frequencies. Note that the transitions $\nu_e \to \bar{\nu}_e$ are prohibited since the diagonal elements of the Majorana neutrino magnetic moments matrix in the flavour basis $\mu^{(f)}$ are zeros due to CPT-invariance. In Fig.4 the probabilities of neutrino oscillations for even higher energy $100$ ZeV are shown. For neutrino energies higher than $100$ ZeV the oscillations probabilities do not depend on the energy. This is similar to the case of two neutrino flavours that was studied in \cite{Kurashvili:2017, Lichkunov:2025rpu}.
In the two flavour case the oscillations probabilities are given by
\begin{eqnarray}\label{p1}
   P_{\nu_e \rightarrow \nu_\mu}(x) &=& \frac{\omega_{vac}^2}{\omega_{vac}^2 + \omega_B^2}\sin^2 2\theta\sin^2\left(\sqrt{\omega_B^2 + \omega^2_{vac}} x\right), \\
   \label{p2}
    P_{\nu_e \rightarrow \bar{\nu}_\mu}(x) &=& \frac{\omega_B^2}{\omega_{vac}^2 + \omega_B^2}\sin^2\left(\sqrt{\omega_B^2 + \omega^2_{vac}}x\right).
\end{eqnarray}
Here $\omega_B = \mu B \cos\alpha$, where $\alpha$ is the Majorana CP-violating phase and $\omega_{vac} = \frac{\Delta m^2}{2p}$. When $\omega_{vac} \ll \omega_B$, which is realized in the case of high neutrino energies, the transition probability $P_{\nu_e \to \nu_\mu}$ vanishes, while $P_{\nu_e \to \bar{\nu}_\mu} = \sin^2\omega_B x$ does not depend on the neutrino energy.

As was shown for the first time in \cite{Popov:2021}, further studied in \cite{Popov:2023wif,Lichkunov:2025rpu} and can be seen from Eqns. (\ref{p1},\ref{p2}), the appearance of the Majorana CP-violating phases alters the picture of neutrino oscillations in a magnetic field. In Fig.5 the probabilities of Majorana neutrino oscillations in the interstellar magnetic field for neutrino energy $100$ ZeV and Majorana CP-violating phases $\alpha_1=\alpha_2=\pi$. One can see that for this particular choice of parameters, the transitions $\nu_e \to \bar{\nu}_\mu$ that were dominant in the absence of the CP-violating phases are now strongly suppressed. In contrast, transitions in the channel $\nu_e \to \bar{\nu}_\mu$ appear with amplitude reaching $0.4$.

\begin{figure}[h]
	\begin{minipage}[h]{0.49\linewidth}
		\center{\includegraphics[width=1\linewidth]{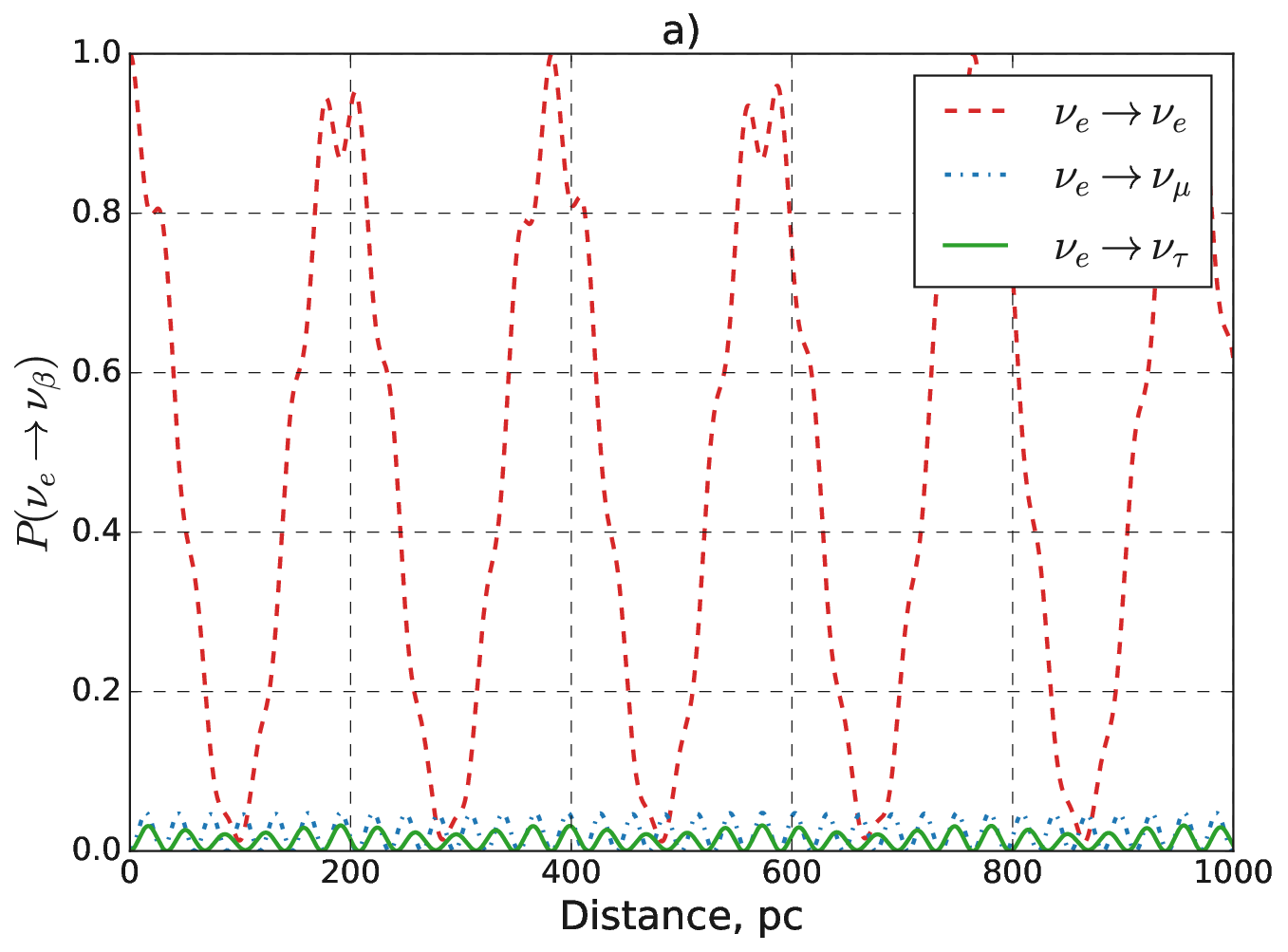}}
	\end{minipage}
	\begin{minipage}[h]{0.49\linewidth}
		\center{\includegraphics[width=1\linewidth]{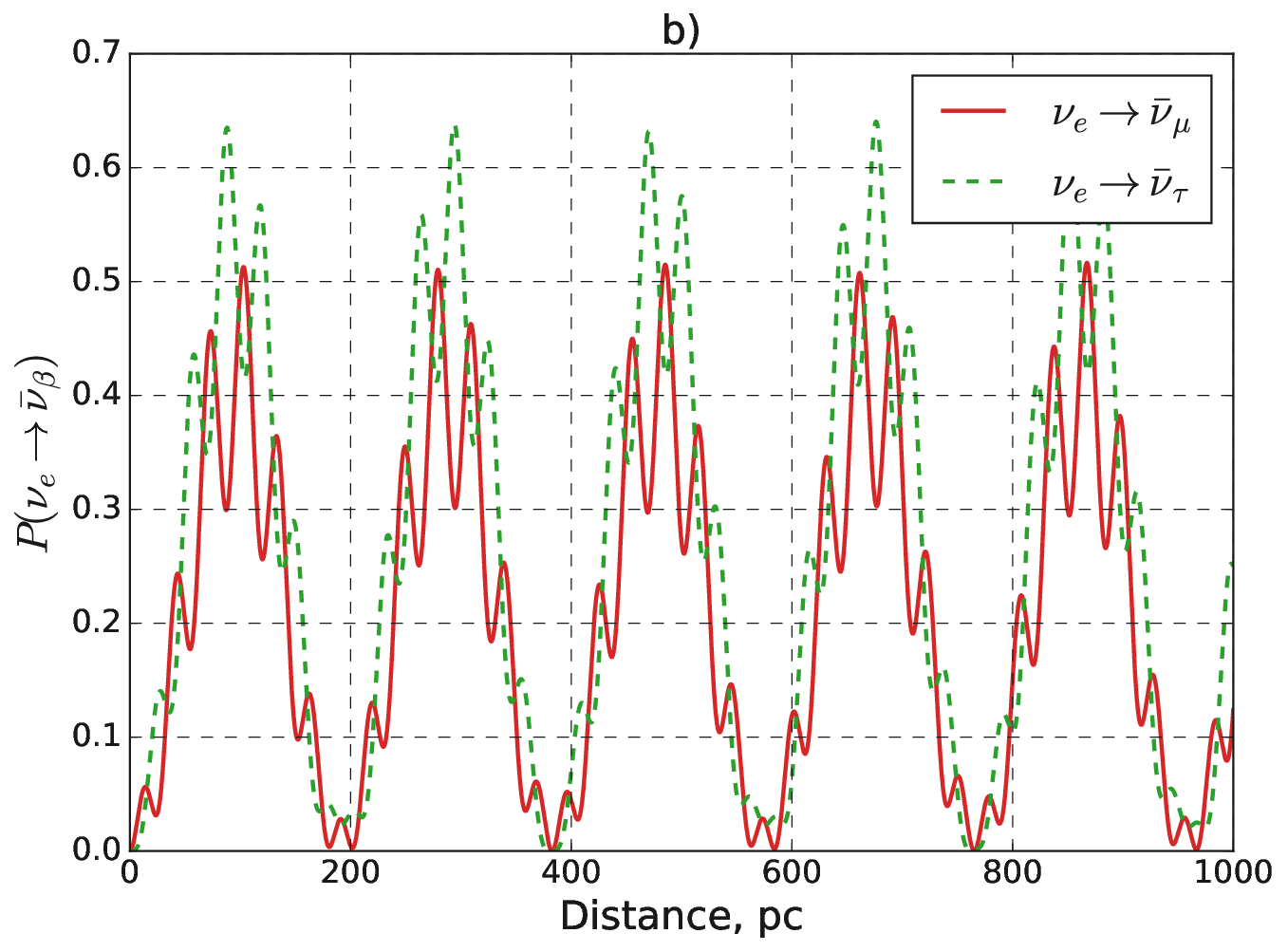}}
	\end{minipage}
	\vspace{-3mm}
	\caption{Majorana neutrino oscillations probabilities for neutrino energy $E = 1$ ZeV and magnetic moments $|\mu_{12}|=|\mu_{13}|=|\mu_{23}| = 6.4\cdot10^{-12} \mu_B$. a) neutrino-neutrino transitions. b) neutrino-antineutrino transitions.}
	\vspace{-1mm}
\end{figure}

\begin{figure}[h]
	\begin{minipage}[h]{0.49\linewidth}
		\center{\includegraphics[width=1\linewidth]{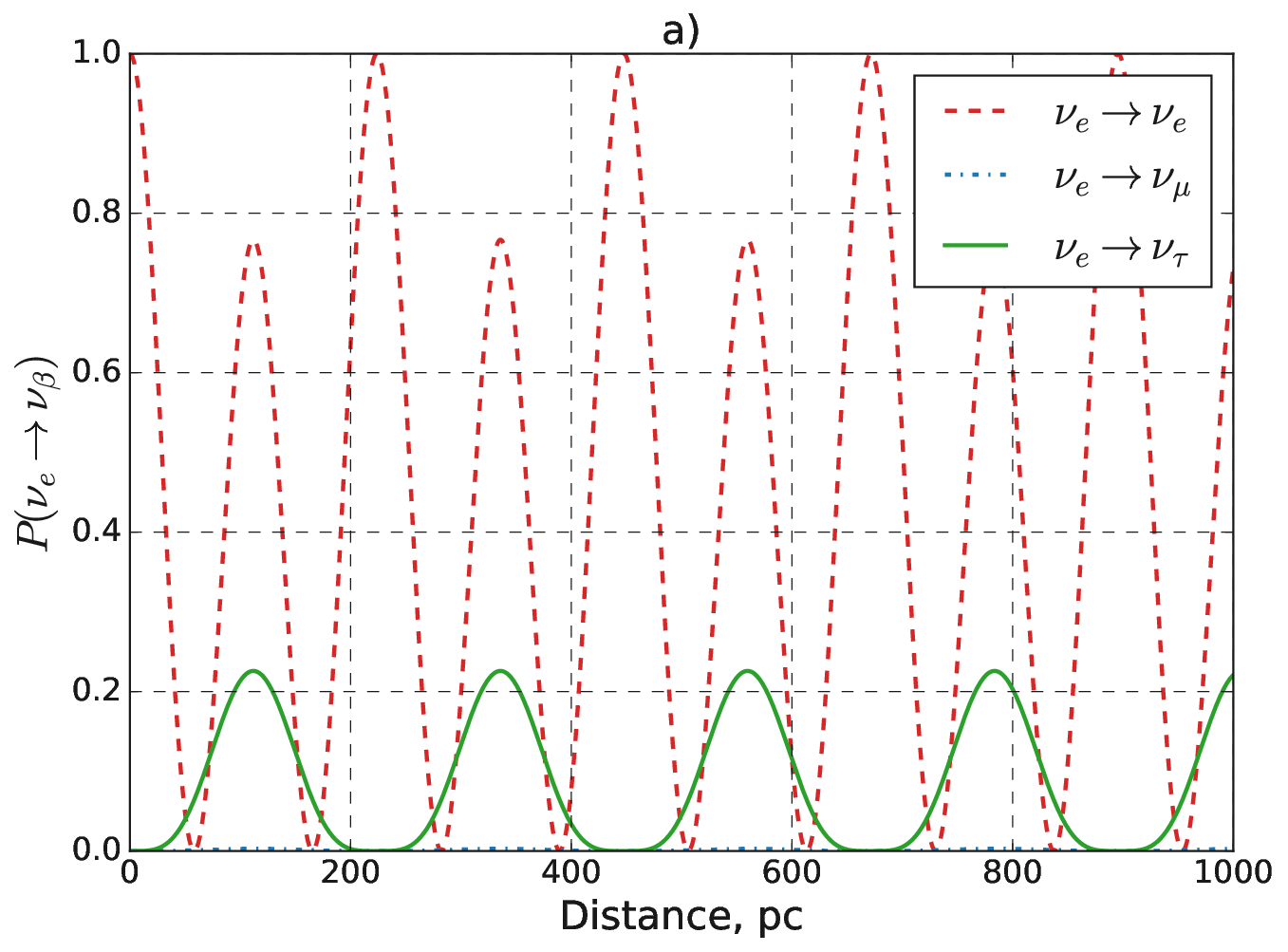}}
	\end{minipage}
	\begin{minipage}[h]{0.49\linewidth}
		\center{\includegraphics[width=1\linewidth]{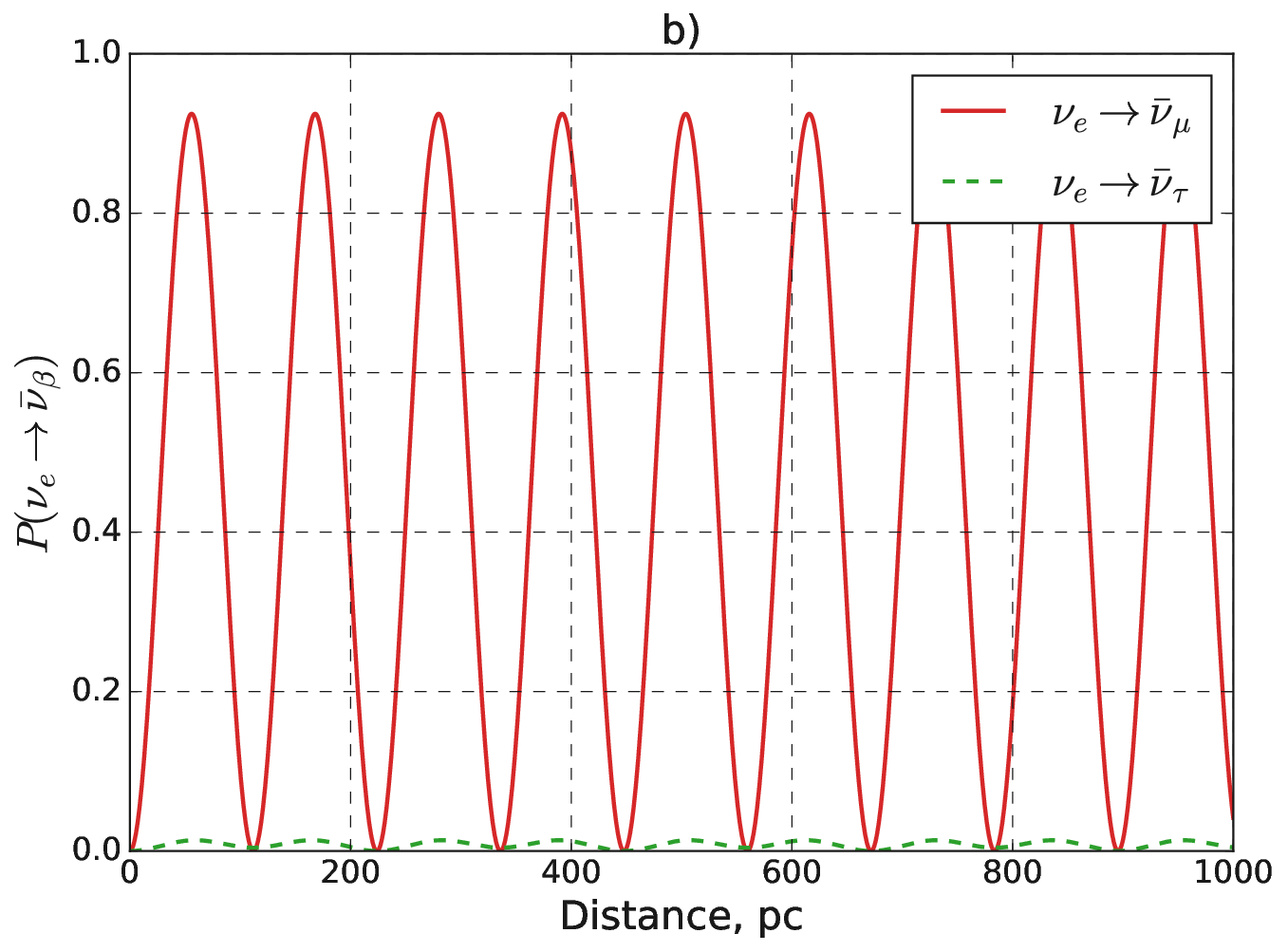}}
	\end{minipage}
	\vspace{-3mm}
	\caption{Majorana neutrino oscillations probabilities for neutrino energy $E = 100$ ZeV and magnetic moments $|\mu_{12}|=|\mu_{13}|=|\mu_{23}| = 6.4\cdot10^{-12} \mu_B$. a) neutrino-neutrino transitions. b) neutrino-antineutrino transitions.}
	\vspace{-1mm}
\end{figure}

\begin{figure}[h]
	\begin{minipage}[h]{0.49\linewidth}
		\center{\includegraphics[width=1\linewidth]{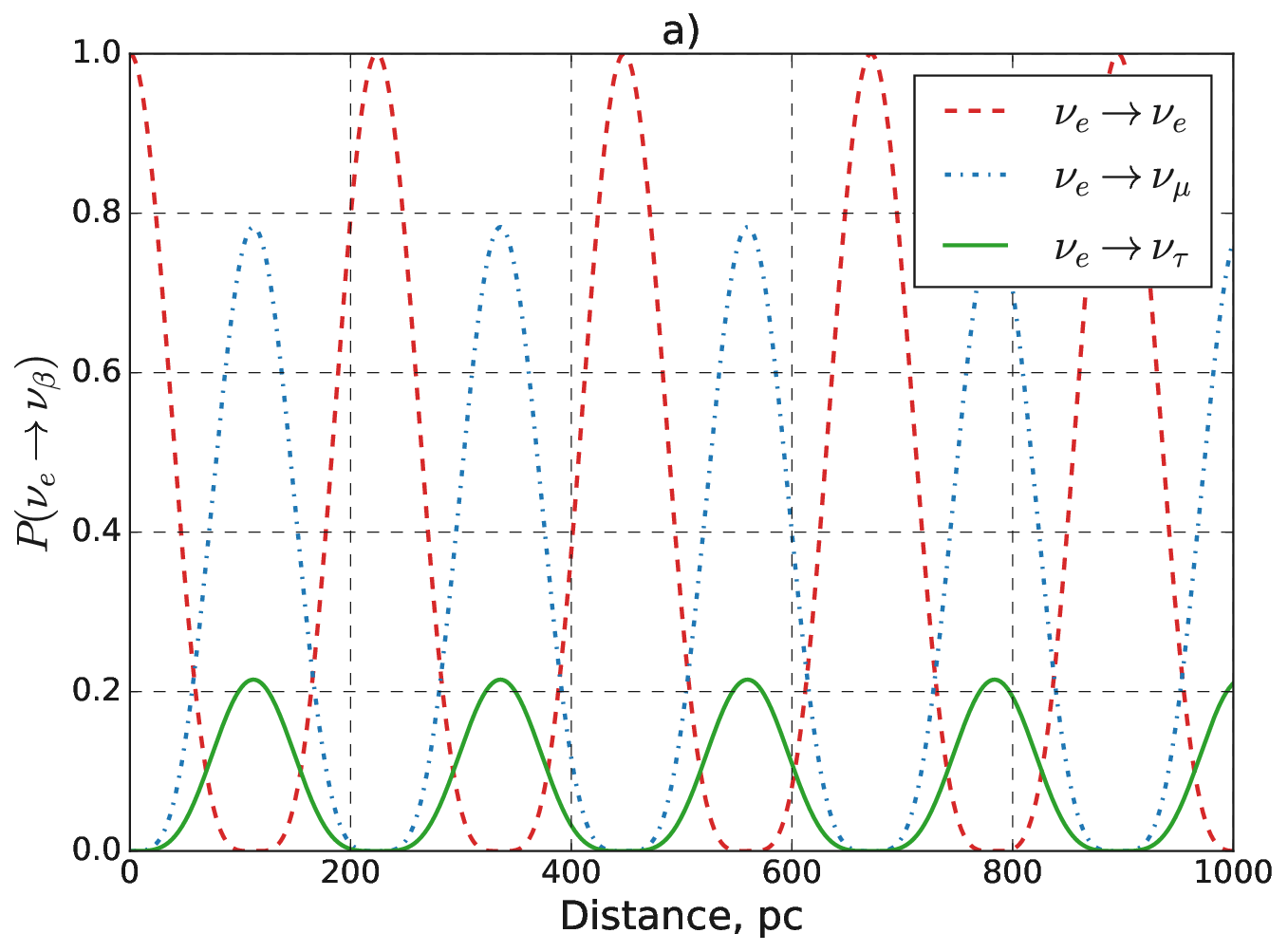}}
	\end{minipage}
	\begin{minipage}[h]{0.49\linewidth}
		\center{\includegraphics[width=1\linewidth]{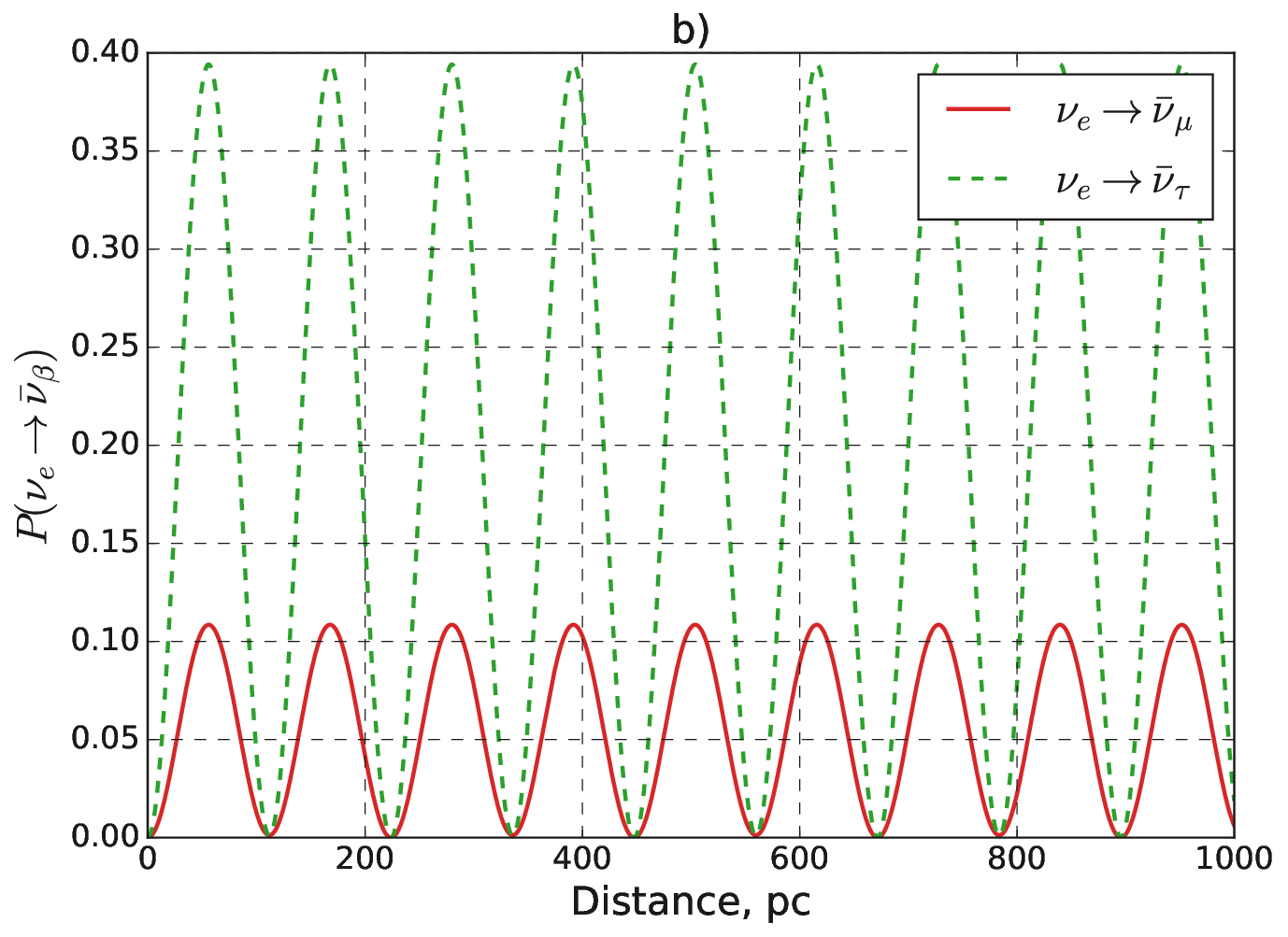}}
	\end{minipage}
	\vspace{-3mm}
	\caption{Majorana neutrino oscillations probabilities for neutrino energy $E = 100$ ZeV and magnetic moments $|\mu_{12}|=|\mu_{13}|=|\mu_{23}| = 6.4\cdot10^{-12} \mu_B$ assuming non-zero Majorana CP-violating phases $\alpha_1=\alpha_2 = \pi$. a) neutrino-neutrino transitions. b) neutrino-antineutrino transitions.}
	\vspace{-1mm}
\end{figure}

\section{Discussion}
Our previous results on neutrino flavour and spin oscillations in magnetic fields \cite{Popov:2019} were extended to the case of three neutrino flavors, as well as to Majorana neutrinos. We have shown that in the Dirac case the probabilities of neutrino flavour oscillations in a magnetic field exhibit a complicated interplay of oscillations on both vacuum $\omega^{vac}_{ij} = \Delta m^2_{ij}/4p$ and magnetic $\omega^B_i = \mu_i B_{\perp}$ frequencies, while spin oscillations probability depends only on the magnetic frequencies $\omega^B_i$. For Majorana case, neutrino-antineutrino transitions mediated by the interaction of the off-diagonal magnetic moments with a magnetic field occur. These transitions are only possible for extreme neutrino energies 10 ZeV and higher. The probabilities of neutrino-neutrino and neutrino-antineutrino transitions depend on the values of the Majorana CP-violating phases. As an example of neutrino oscillations in a magnetic field we have considered oscillations of ultra-high energy neutrinos in the interstellar magnetic field.

Our findings on three flavour neutrino oscillations show that the presence of interstellar magnetic field indeed can modify neutrino flux coming from a point source and observed by terrestrial neutrino telescopes. From (\ref{lengths_B}) it follows that the oscillations due to interaction with the magnetic field on the Galactic scale can occur for neutrino magnetic moments $\sim 10^{-13} \mu_B$ and higher, which is almost an order of magnitude lower than the best current upper bounds. Furthermore, there is evidence of the observation of a magnetic field of strength $\sim 1 \mu$G on the megaparsec scale in the Local Supercluster \cite{Valee:2002}. Thus, for extragalactic neutrino source interaction with a magnetic field can significantly modify flavour composition even for neutrino magnetic moments as small as $10^{-15} \mu_B$.

It is predicted that high-energy and ultra-high energy neutrinos flavour composition at source follows pattern $\Phi_{e}^0 : \Phi_{\mu}^0 : \Phi_{\tau}^0 \approx 1:2:0$ \cite{Katz:2011ke}. Then, for the case of vacuum oscillations, neutrino flavour composition at the terrestrial neutrino telescope is $(r_{e}^{vac}, r_{\mu}^{vac}, r_{\tau}^{vac}) \approx (1/3, 1/3, 1/3)$ \cite{Xing:2009zzb}.

Using the calculated three flavour oscillations probabilities, one can compute fluxes of neutrinos at the distance $x$ from neutrino source accounting their interaction with a magnetic field. Since current and forthcoming neutrino telescopes can not distinguish neutrinos and antineutrinos (with a rare exception of detection via Glashow resonance), we compute total flux of neutrinos and antineutrinos of flavour $\alpha$. For Dirac neutrinos we obtain
\begin{equation}\label{flux_dirac}
	\Phi_{\alpha}^D(x) = \sum_{\beta} \Phi^0_{\beta} P_{\nu^L_{\beta} \to \nu^L_{\alpha}}(x),
\end{equation}
where $\Phi^0_{\beta}$ are the total neutrino and antineutrino fluxes at the source, and calculate neutrino flavour composition observed by a neutrino telescope:
\begin{equation}\label{flavour_composition_dirac}
	r_\alpha^D(x) = \frac{\Phi^D_\alpha(x)}{\sum_\beta \Phi^D_\beta (x)},
\end{equation}
where $\alpha,\beta = e,\mu,\tau$. Eqns. (\ref{flux_dirac}) and (\ref{flavour_composition_dirac}) were derived using $P_{\nu^L_{\beta} \to \nu^L_{\alpha}} \approx P_{\bar{\nu^L_{\beta}} \to \bar{\nu^L_{\alpha}}}$ (see \ref{prob_fl_final}). A detailed analysis of the Galactic magnetic field effects on the high-energy and ultra-high energy Dirac neutrinos flavour composition is given in \cite{Popov:2024spe}.

For Majorana neutrinos, to calculate the flavour composition we also need to account for the neutrino-antineutrinos transitions.

\begin{equation}\label{flavour_composition_majorana}
    r_\alpha^M(x) = \sum_\beta r^0_\beta (P_{\nu_\beta \to \bar{\nu}_\alpha}(x) + P_{\nu_\beta \to \nu_\alpha}(x)) + \sum_{\beta} \bar{r}^0_\beta (P_{\bar{\nu}_{\beta} \to \bar{\nu}_\alpha}(x) + P_{\bar{\nu}_{\beta} \to \nu_\alpha}(x)),
\end{equation}
where $\beta = e,\mu,\tau$, and $r^0_\alpha$ and $\bar{r}^0_\alpha$ are the ratios of neutrino and antineutrinos of flavour $\alpha$ at the source correspondingly. However, for Majorana neutrinos effects due to magnetic fields become pronounced only for extremely high energies and are unlikely to be observed in the nearest future.

Consider Dirac neutrino transformation into sterile state under such conditions. As it is seen from eq. (\ref{spin_total}), in case of the neutrino magnetic moments equality ($\mu_1=\mu_2=\mu_3$) the probability of the spin conversation
\begin{equation}
P_{\nu_e^L \to \nu^R}(x) = \cos^2(\mu Bx),
\end{equation}

It is also interesting to consider the magnetic field effects on the diffuse high-energy and ultra-high energy neutrino fluxes. Assuming that left-handed (active) neutrinos are born inside the sphere of radius $L$ around the Earth, we can estimate the fraction of left-handed neutrinos $Y_{\nu^L}$ that reach a terrestrial detector as follows
\begin{equation}
    Y_{\nu^L}=\frac{1}{L}\int_{-L}^0\cos^2(\mu Bx)dx.
\end{equation}

After integrating, we obtain the following expression
\begin{equation}
    Y_{\nu^L}= \frac{1}{2} + \frac{\sin(2\mu BL)}{4\mu BL}.
\end{equation}

As seen, in the case $\mu B L \gg 1$ the fraction of left-handed neutrinos approximately equals 0.5, and interaction with a magnetic field leads to the reduction of the overall neutrinos flux. The flavour composition of the diffuse neutrino flux is unaffected by the interaction with a magnetic field since the flavour oscillations probabilities (\ref{prob_fl}) are averaged out to $P_{\nu_{\alpha}^L \to \nu_{\beta}^L} = \sum_{i=1}^3 |U_{\alpha i}|^2 |U_{\beta i}|^2 $ due to integration over distance.

It is likely that in next 20 years neutrino telescopes such as Baikal-GVD, KM3NeT, P-ONE, TAMBO and IceCube-Gen2 will collect enough data to determine the flavour composition of high-energy cosmic neutrino fluxes \cite{Song:2020nfh}. If the measured composition contradicts the predicted pattern (1/3, 1/3, 1/3), this might be a signal of new physics, including neutrino magnetic moments. Using the probabilities of neutrino oscillations in interstellar magnetic field we can study these possible effects of nonzero neutrino magnetic moments on UHE neutrino fluxes flavour composition.

The results are also of interest for experiments that are sensitive to the electromagnetic scattering of astrophysical neutrinos \cite{Kouzakov:2024xnq,Kouzakov:2025qcf}, including the future SATURNE experiment \cite{Cadeddu:2024vzt}, in which Solar neutrino fluxes are likely to be detected.

\section*{Acknowledgements}
The work is supported by the Russian Science Foundation under grant No.24-12-00084.


\end{document}